\documentclass{article}
\usepackage{spconf,amsmath,graphicx}
\usepackage{booktabs}
\usepackage{amsfonts}
\usepackage{multirow}
\usepackage{bbding}
\usepackage{pifont}
\title{MARGIN-MIXUP: A METHOD FOR ROBUST SPEAKER VERIFICATION IN MULTI-SPEAKER AUDIO}
\name{Jenthe Thienpondt, Nilesh Madhu\thanks{This work is supported by the Research Foundation - Flanders (FWO) under grant numbers G081420N and S004923N.}, Kris Demuynck}
\address{IDLab, Department of Electronics and Information Systems, Ghent University - imec, Belgium\\
{jenthe.thienpondt@ugent.be, nilesh.madhu@ugent.be, kris.demuynck@ugent.be}}
\begin{document}
% \ninept
%
\maketitle
%
% 100 to 150 words! Current: 136
% TODO:
% update final performance number
\begin{abstract}
This paper is concerned with the task of speaker verification on audio with multiple overlapping speakers. Most speaker verification systems are designed with the assumption of a single speaker being present in a given audio segment. However, in a real-world setting this assumption does not always hold. In this paper, we demonstrate that current speaker verification systems are not robust against audio with noticeable speaker overlap. To alleviate this issue, we propose margin-mixup, a simple training strategy that can easily be adopted by existing speaker verification pipelines to make the resulting speaker embeddings robust against multi-speaker audio. In contrast to other methods, margin-mixup requires no alterations to regular speaker verification architectures, while attaining better results. On our multi-speaker test set based on VoxCeleb1, the proposed margin-mixup strategy improves the EER on average with 44.4\% relative to our state-of-the-art speaker verification baseline systems.
\end{abstract}
\begin{keywords}
speaker verification, multiple speakers, mixup, ECAPA-TDNN, ResNet
\end{keywords}

% TODO:
% - link to figure somewhere
% - maybe more exact explanation of multi-speaker speaker verification 
% - maybe a real-world example of the usage of multi-speaker speaker verification?
\section{Introduction}
\label{sec:introduction}
Speaker verification tries to answer the question if two utterances originate from the same person. Speaker verification systems have recently gained impressive results due to the abundance of labelled training data \cite{vox1, vox2} and the advent of deep learning based speaker recognition architectures~\cite{ecapa_tdnn, freq_paper}.

Most speaker recognition pipelines are based on the assumption of a single speaker being present in training and test utterances. Although augmentation techniques such as babble noise~\cite{musan} are common to provide a certain robustness towards background speakers, we argue that these systems fail to distinguish speakers in a multi-speaker audio scenario.

In prior work done on speaker verification in multi-speaker audio, a few approaches have been suggested. The most straightforward method is the usage of speaker diarization prior to speaker verification~\cite{speaker_overlap_diarization}. These systems perform well in the case of limited temporal overlap between speaker turns but are not designed to handle mixtures with significant interference between speakers. In~\cite{multi_speaker_extraction}, target-speaker extraction~\cite{universal_speaker_extraction} is applied in a joint training scheme with speaker verification to handle the significant speaker overlap. Another approach is used in~\cite{multi_speaker_temp_fusion}, with an architecture injecting the target speaker embedding into the hidden frame-level features before the statistics pooling layer. However, these approaches use task-specific architectures or require significant adaptations of existing speaker verification pipelines.

To alleviate this issue, we propose margin-mixup, a training strategy based on mixup \cite{mixup} to produce speaker embeddings significantly more robust in the multi-speaker scenario. Margin-mixup mixes input waveforms on the batch-level in combination with the usage of an adapted version of the Additive Angular Margin (AAM)~\cite{arcface} softmax loss function, used in current state-of-the-art speaker verification systems~\cite{ecapa_tdnn, score_shift}. The margin-mixup training strategy can be employed by existing speaker verification models without any computational overhead and requires no architectural alterations.

The paper is organized as follows: Section 2 gives an overview of current state-of-the-art speaker verification systems and our baseline models. Section 3 describes the proposed margin-mixup training strategy. Subsequently, Section 4 explains the experimental setup to verify the proposed margin-mixup strategy while Section 5 analyzes the results. Finally, Section 6 gives some concluding remarks.

% The paper is organized as follows: Section 2 gives an overview of current state-of-the-art speaker verification systems and the baseline models used in this paper. Section 3 describes the proposed margin-mixup training strategy. Subsequently, Section 4 explains the experimental setup to verify the effectiveness of the proposed margin-mixup training strategy while Section 5 analyzes the consequent results. Finally, Section 6 gives some concluding remarks.

% The paper is organized as follows: Section 2 gives an overview of current state-of-the-art speaker verification and baseline systems used in this paper. Section 3 describes the proposed margin-mixup training strategy. Section 4 explains the experimental setup to verify the effectiveness of the proposed margin-mixup training strategy while Section 5 analyzes the consequent results. Finally, Section 6 gives some concluding remarks.

\section{Baseline systems}
\label{sec:baseline_system}
We verify the multi-speaker verification performance of three state-of-the-art models and use those as our baseline for the proposed margin-mixup training strategy. Our first baseline system is ECAPA-TDNN~\cite{ecapa_tdnn}, which improves upon the x-vector~\cite{x_vectors} architecture with an adapted version of Squeeze-Excitation (SE)~\cite{se_block}, channel- and context dependent statistics pooling and the aggregation of multi-layer features. Secondly, we employ the ECAPA-CNN-TDNN model \cite{freq_paper}, a hybrid architecture which adds a 2D-convolutional stem to the ECAPA model to enhance its capability to model frequency independent features. Lastly, we consider fwSE-ResNet34~\cite{score_shift}, a ResNet based architecture with an adapted frequency-wise SE layer and usage of positional encodings.

%Current peaker verification systems consist of either enhanced Time Delay Neural Networks (TDNN)~\cite{tdnn} or ResNet~\cite{resnet} based architectures.

% TODO
% - refer to recent speaker verification competitions for first claim
\section{Margin-mixup}
\label{sec:margin_mixup}
Current state-of-the-art speaker verification pipelines attain impressive performance on audio containing a single speaker. While those systems are generally robust against babble-like background noise, we argue they are not able to handle significant interference of multiple speakers. This is mainly due to the emphasis of single-speaker audio in most speaker recognition training datasets and the subsequent absence of a suitable loss function to model speaker overlap. This is further accentuated by the usage of margin based loss functions, such as the commonly used AAM-softmax. Margin based loss functions increase discriminative performance of speaker recognition systems by inducing a margin penalty on a single target speaker during training. However, this makes margin-based loss functions unsuitable for modelling multiple overlapping speakers.

To alleviate this issue, we introduce margin-mixup, a training strategy which enables the speaker embedding space to model overlapping speakers. As we demonstrate, margin-mixup can easily be adopted by regular speaker verification models without any alterations to the speaker embedding extractor architecture, making it a more flexible method compared to task-specific multi-speaker verification approaches.

% \subsection{Mixup training for speaker recognition}
\subsection{Proposed margin-mixup training strategy}
\label{ssec:mixup_training}
Margin-mixup is based on the mixup training strategy introduced in \cite{mixup}, which proposes to sample input features during training from a vicinal distribution by the linear interpolation of input features. With the consequent interpolation of target labels, the model extends the embedding space with a notion of in-between classes.

In this paper, we keep the interpolation limited to two speakers. To adapt the mixup training strategy for speaker recognition systems, an interpolated input waveform $\hat{\mathbf{x}}$ is constructed as following:

\begin{equation}
\label{eq:mixup_x}
\hat{\mathbf{x}} = \lambda \frac{\mathbf{x}_{a}}{\lVert \mathbf{x}_{a} \rVert} + (1-\lambda) \frac{\mathbf{x}_{b}}{\lVert \mathbf{x}_{b} \rVert}
\end{equation}

with $\lambda$ being the interpolation strength between two single-speaker waveforms $\mathbf{x}_{a}$ and $\mathbf{x}_{b}$. The input waveforms are energy normalized before the weighted interpolation. Subsequently, the corresponding interpolated target label $\mathbf{\hat{y}}$ is defined as:

\begin{equation}
\label{eq:mixup_y}
\mathbf{\hat{y}} = \lambda \mathbf{y}_a + (1-\lambda) \mathbf{y}_b
\end{equation}

with $\mathbf{y}_a$ and $\mathbf{y}_b$ being one-hot label vectors. We note that the initial intended goal of mixup training is to make systems generalize better on the unmixed task and making them less prone to corrupt labels \cite{mixup}. However, our main goal during mixup training is to learn an embedding space which can cope with overlapping speakers. 

% cite some examples of other speaker ID systems using AAM after claim
% \subsection{Proposed margin-mixup}
\subsection{Margin penalty mixing}
\label{ssec:margin_mixup}
The interpolated target label $\mathbf{\hat{y}}$ poses a problem for margin based loss functions as they are designed to impose a margin penalty on a single target label. Since current state-of-the-art speaker verification systems are based on such loss functions, we use an adapted version of AAM-softmax in our proposed margin-mixup training strategy.

AAM-softmax is based on the cosine distance between a speaker embedding $\mathbf{e}_i$ with speaker label $i$ and the corresponding class center $\pmb{W}_{i}$ with $\pmb{W} \in \mathbb{R}^{D \times N}$. $D$ and $N$ indicate the embedding size and number of training speakers, respectively. In addition, a margin penalty is applied on the angle between the speaker embedding and the target class to enforce a tighter boundary around the speaker classes and subsequently improve speaker verification performance.

% notation must be more clear, no consistency of theta in equations
During margin-mixup training, the margin penalty $m$ is applied on the angle $\theta_{i}$ between the embedding of the mixed input utterance $\mathbf{e}_{a,b}$ containing both speakers $a$ and $b$ and their corresponding class centers $\pmb{W}_{i}$ when $i \in [a,b]$ weighted according to the interpolation value $\lambda$:

\begin{equation}
\label{eq:margin_mixup_angle}
    \hat{\theta}_{i}=
    \begin{cases}
      \theta_{i} + \lambda m , & \text{if}\ i=a \\
      \theta_{i} + (1-\lambda) m , & \text{if}\ i=b \\
      \theta_{i}, & \text{else}
    \end{cases}
\end{equation}

% \begin{equation}
% \label{eq:margin_mixup_angle}
%     \hat{\theta}_{a, b, t}=
%     \begin{cases}
%       \theta_{a, b, t} + \lambda m , & \text{if}\ a=t \\
%       \theta_{a, b, t} + (1-\lambda) m , & \text{if}\ b=t \\
%       \theta_{a, b, t}, & \text{else}
%     \end{cases}
% \end{equation}

With the applied margin penalty given in Equation \ref{eq:margin_mixup_angle}, the margin-mixup loss function $L$ is subsequently given by:

\begin{equation}
\label{eq:margin_mixup_loss}
L = \lambda \log \frac{e^{s\:cos(\hat{\theta}_{a})}}{\sum_{j=1}^N e^{s\:cos(\hat{\theta}_{j})}} + (1-\lambda) \log \frac{e^{s\:cos(\hat{\theta}_{b})}}{\sum_{j=1}^N e^{s\:cos(\hat{\theta}_{j})}}
\end{equation}

with $\hat{\theta}_{a}$ and $\hat{\theta}_{b}$ indicating the angle with the applied margin penalty between the multi-speaker input embedding $\mathbf{e}_{a,b}$ and the class centers of $a$ and $b$, respectively. The hyperparameter $s$ is a scale factor to optimize the gradient flow during training.

% maybe create a plot here which links lambda to SNR ratio?
We follow the original paper on mixup \cite{mixup} to determine the interpolation weight $\lambda$ by sampling the value from a beta distribution with $\alpha$ and $\beta$ equal to 0.2. This will ensure the interpolation of features is mostly focused on one of the utterances, as we do not want to impact the performance on regular speaker verification with single-speaker test utterances.

% \begin{figure}[htb]
% \begin{minipage}[b]{.48\linewidth}
%   \centering
%   \centerline{\includegraphics[angle=-90, width=4.3cm]{tsne_normal.png}}
% %  \vspace{1.5cm}
%   \centerline{No margin-mixup}\medskip
% \end{minipage}
% \hfill
% \begin{minipage}[b]{0.48\linewidth}
%   \centering
%   \centerline{\includegraphics[width=4.3cm]{tsne_mm.png}}
% %  \vspace{1.5cm}
%   \centerline{With margin-mixup}\medskip
% \end{minipage}
% %
% \caption{UMAP reduced embeddings of utterances from two speakers mixed together with various SNR levels from a model trained without (left) and with (right) margin-mixup. Notice that the embedding space in the model trained with margin-mixup has a better notion of in-between speakers.}
% \label{fig:res}
% %
% \end{figure}

\section{Experimental setup}
\label{sec:experimental_setup}

% TODO
% recheck italizing datasets
\subsection{Multi-speaker test dataset}
To analyze the proposed margin-mixup training strategy for multi-speaker audio settings, we construct a custom test set Vox1-M based on the original VoxCeleb1~\cite{vox1} (Vox1-O) test set. The overlapping speaker set Vox1-M is constructed by the addition of an interfering utterance from the unused training partition of VoxCeleb1 to the Vox1-O utterances. The corresponding signal-to-noise ratio~(SNR) is uniformly sampled between 0-5dB. If necessary, we repeat the utterance of the interfering speaker to fully overlap the target speaker utterance. The resulting multi-speaker test set is challenging due to the in-domain utterance mixing combined with low SNR levels. All models are trained using the training partition of VoxCeleb2~\cite{vox2} and will be evaluated by the EER metric.

% \subsection{Speaker embedding extractors}
% Our ECAPA-TDNN baseline model follows the same structure as defined in~\cite{ecapa_tdnn} with a hidden feature dimension of 1024 and an additional SE-Res2Block with kernel size 3 and dilation factor 5 at the end of the frame-level feature extractor. The fwSE-ResNet34 and ECAPA-CNN-TDNN models correspond to the architectures presented in~\cite{freq_paper}. More details about these models can be found in the accompanying papers.

\subsection{Training configuration}
Our ECAPA-TDNN baseline model follows the same structure as defined in~\cite{ecapa_tdnn} with a hidden feature dimension of 1024 and an additional SE-Res2Block with kernel size 3 and dilation factor 5 at the end of the frame-level feature extractor. The fwSE-ResNet34 and ECAPA-CNN-TDNN models correspond to the architectures presented in~\cite{freq_paper}. More details about these models can be found in the accompanying papers.

During training, we take random crops of 2 seconds of audio to prevent overfitting. In addition, we apply one random augmentation based on the MUSAN library~\cite{musan} (additive noise, music and babble) or the RIR corpus \cite{rirs} (reverb). Our input features consists of 80-dimensional log Mel-filterbanks with a window length and hop length of 25ms and 10ms, respectively. Subsequently, we apply SpecAugment~\cite{specaugment} which randomly masks 0-10 contiguous frequency bins and 0-5 time frames. Finally, we mean normalize the log filterbank energies across the temporal dimension. The margin penalty $m$ and scale factor $s$ of the AAM-softmax loss function are set to 0.2 and 30, respectively.

To optimize the speaker embedding extractors, we use the Adam~\cite{adam} optimizer with a weight decay of 2e-4. A Cyclical Learning Rate (CLR)~\cite{clr} strategy is used with the \textit{triangular2} decaying strategy and a minimum and maximum learning rate of 1e-8 and 1e-3, respectively. The cycle length is set to 130K iterations with a batch size of 128.

We apply large margin fine-tuning (LM-FT) \cite{icassp_voxsrc20} after the initial training phase to encourage intra-speaker compactness and increase inter-speaker distances. The margin penalty and crop length are raised to 0.5 and 5s, respectively. All augmentations are disabled during fine-tuning. The CLR maximum learning rate is decreased to 1e-5 with the cycle length lowered to 60K.

% \subsection{Score normalization}
After training the speaker embedding extractor, we score the test trials by computing the cosine distance between the enrollment and target speaker embeddings. Subsequently, we apply top-1000 adaptive s-normalization \cite{s_norm, s_norm_2} with an imposter cohort consisting of the average speaker embedding of the training utterances in VoxCeleb2.

% TODO
% explain benefits of batch-level mixing here or above
\subsection{Margin-mixup configuration}
We apply the proposed margin-mixup training strategy described in Section \ref{sec:margin_mixup} during both the initial and fine-tuning training phase with the interpolation weight $\lambda$ sampled from a beta distribution with both $\alpha$ and $\beta$ set to 0.2. %To increase computational efficiency, the mixing of utterances is applied on the batch-level.

% % TODO
% % fix alignment of figures
% \begin{figure}[h]
% \begin{minipage}[b]{.49\linewidth}
%   \centering
% %   \centerline{\includegraphics[width=4.3cm]{tsne_standard_v2.png}}
%   \centerline{\includegraphics[width=3.48cm]{images/no_mm_full_long_no_label.png}}
% %   \centerline{\includegraphics[width=4.6cm]{images/no_mm_full_long_no_cb.png}}
% %  \vspace{1.5cm}
% %   \centerline{No margin-mixup}\medskip
% \end{minipage}
% \hfill
% \begin{minipage}[b]{0.49\linewidth}
%   \centering
% %   \centerline{\includegraphics[width=4.3cm]{tsne_mm_v2.png}}
%   \centerline{\includegraphics[width=4.6cm]{images/with_mm_full_long.png}}
% %   \centerline{\includegraphics[width=4.6cm]{images/with_mm_full_long.png}}
% %  \vspace{1.5cm}
% %   \centerline{With margin-mixup}\medskip
% \end{minipage}
% %
% \caption{UMAP-reduced~\cite{umap} embeddings of utterances from two speakers, indicated by the circle and cross marker, mixed together with various SNR levels from a model trained without (left) and with (right) margin-mixup. Notice that the embedding space in the model trained with margin-mixup has a better notion of in-between speakers.}
% \label{fig:res}
% %
% \end{figure}

\begin{figure}[h]
\begin{minipage}[b]{1\linewidth}
  \centering
  \centerline{\includegraphics[width=9.2cm]{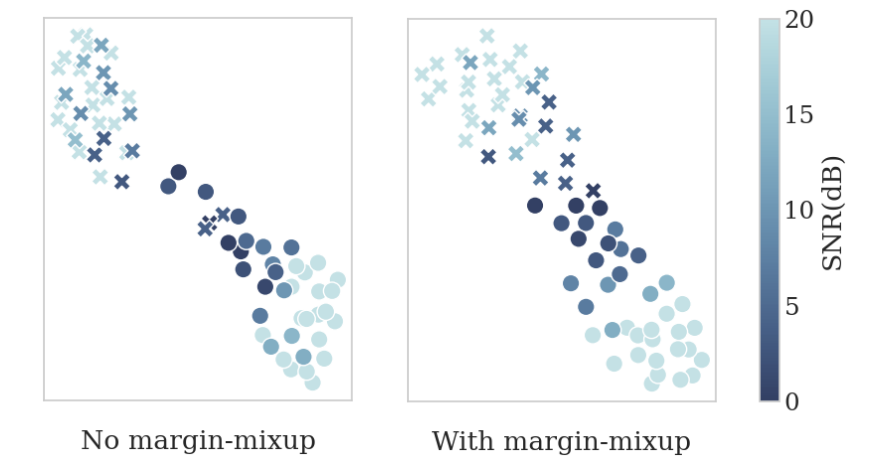}}
\end{minipage}
\caption{UMAP-reduced~\cite{umap} embeddings of utterances from two speakers, indicated by the circle and cross marker, mixed together with various SNR levels from a model trained without (left) and with (right) margin-mixup. Notice that the embedding space in the model trained with margin-mixup has a better notion of in-between speakers.}
\label{fig:res}
\end{figure}

\begin{table}[ht]
  \centering
  \begin{tabular}{lccc}
    \toprule
      \textbf{Systems} & \textbf{M-M} & \multicolumn{2}{c}{\textbf{EER(\%)}} \\
      \cmidrule(lr){3-4}
      & & \multicolumn{1}{c}{\textbf{Vox1-O}} & \multicolumn{1}{c}{\textbf{Vox1-M}} \\
    \midrule
    \multirow{2}{*}{ECAPA-TDNN} & - & 0.76 & 22.31 \\
     & \checkmark & \textbf{0.74} & \textbf{12.42} \\
    \midrule
    \multirow{2}{*}{ECAPA-CNN-TDNN} & - & \textbf{0.66} & 22.15 \\
     & \checkmark & 0.68 & \textbf{12.13} \\
    \midrule
    \multirow{2}{*}{fwSE-ResNet34} & - & \textbf{0.63} & 21.11 \\
     & \checkmark & 0.66 & \textbf{11.85} \\
     \bottomrule
  \end{tabular}
    \caption{Performance analysis of the margin-mixup~(M-M) strategy on the Vox1-O and multi-speaker Vox1-M test sets.}
    \label{tab:performance_mixed_margin}
\end{table}

% TODO: 
% change baseline model to resnet in text
% recalc all relative performance
\section{Results}
\label{sec:results}

Table \ref{tab:performance_mixed_margin} gives an overview of the impact of applying the proposed margin-mixup strategy on the baseline models described in Section \ref{sec:baseline_system}. We notice a significant performance degradation across all baseline models when mixing another speaker in the test trials as done in the Vox1-M test set. This supports our hypothesis that the single-speaker assumption held in most speaker verification setups results in embeddings unable to model significant speaker overlap. However, while margin-mixup has a negligible performance impact on the single-speaker test set Vox1-O, the EER on Vox1-M improves on average with 44.4\% relative over the baseline systems. This shows that the proposed margin-mixup training strategy significantly helps to create an embedding space which successfully models in-between speakers, as illustrated in Figure \ref{fig:res}, without impacting performance on the single-speaker scenario.

\begin{table}[ht]
  \centering
  \begin{tabular}{llcc}
    \toprule
     & \textbf{Method} & \multicolumn{1}{c}{\textbf{Vox1-O}} & \multicolumn{1}{c}{\textbf{Vox1-M}} \\
    \midrule
    & fwSE-ResNet34 & 0.66 & 11.85 \\
    \midrule
    \midrule
    A & no mixed margin & 0.77 & 13.15 \\
    B & no mixup loss & 0.74 & 17.23 \\
    C & only input mixup & 0.78 & 17.65 \\
     \bottomrule
  \end{tabular}
    \caption{Ablation study on the components of margin-mixup.}
    \label{tab:ablation}
\end{table}

To get a better understanding of the impact of the various components of the margin-mixup training strategy, an ablation study is done in Table \ref{tab:ablation}. The baseline system is the fwSE-ResNet34 model trained with margin-mixup. Note that all ablation experiments in this table are still using the input mixup as described in Section \ref{ssec:mixup_training}. In experiment \textit{A}, we trained the model without mixing the AAM-softmax margins and applied the margin penalty only to the original speaker by setting $\lambda = 1$ in Equation \ref{eq:margin_mixup_angle}. Without mixing the margins, a performance degradation on both Vox1-O and Vox1-M are observed, indicating the importance of properly mixed margin penalties to fully exploit the AAM-softmax loss function in a multi-speaker setup. When not applying the mixup loss by setting $\lambda = 1$ in Equation \ref{eq:margin_mixup_loss} and consequently not imposing the model to explicitly detect both speakers, we see a large performance degradation on Vox1-M of 45.4\% EER relative. Experiment \textit{C} takes this further by setting $\lambda = 1$ in both Equation \ref{eq:margin_mixup_angle} and \ref{eq:margin_mixup_loss}, with an additional degradation on both test sets. Notably, the performance on Vox1-M is still an improvement over the fwSE-ResNet34 baseline model without margin-mixup in Table \ref{tab:performance_mixed_margin}. We suspect this is due to the input still being mixed with another speaker and can be regarded as a regular augmentation, bringing the training condition closer to the multi-speaker test setup of Vox1-M.

\begin{table}[ht]
  \centering
  \begin{tabular}{lcccc}
    \toprule
     \textbf{Systems} & \multicolumn{1}{c}{\textbf{Vox1-O}} & & \multicolumn{1}{c}{\textbf{Vox1-M}} & \\
     \cmidrule(lr){3-5}
      & & \textbf{0dB} & \textbf{2dB} & \textbf{0-5dB}\\
    \midrule
    baseline & \textbf{0.63} & 33.67 & 23.11 & 21.11\\
    \midrule
    \midrule
    $\alpha, \beta = 0.1$ & 0.64 & 19.12 & 13.12 & 12.82\\
    $\alpha, \beta = 0.2$ & 0.66 & 18.64 & 12.57 & 11.85\\
    $\alpha, \beta = 0.4$ & 0.74 & 17.32 & 11.63 & 11.12\\
    $\alpha, \beta = 0.8$ & 0.94 & 15.23 & 10.62 & 10.31\\
    $\alpha, \beta = 1$ & 1.12 & \textbf{14.84} & \textbf{9.44} & \textbf{8.93}\\
     \bottomrule
  \end{tabular}
    \caption{Analysis of the impact of different $\alpha$ and $\beta$ parameters in the beta distribution used to sample the interpolation weight $\lambda$ in the margin-mixup strategy.}
    \label{tab:beta_distribution}
\end{table}

\begin{figure}[ht]
\begin{minipage}[h]{1.0\linewidth}
  \centering
  \centerline{\includegraphics[scale=0.33]{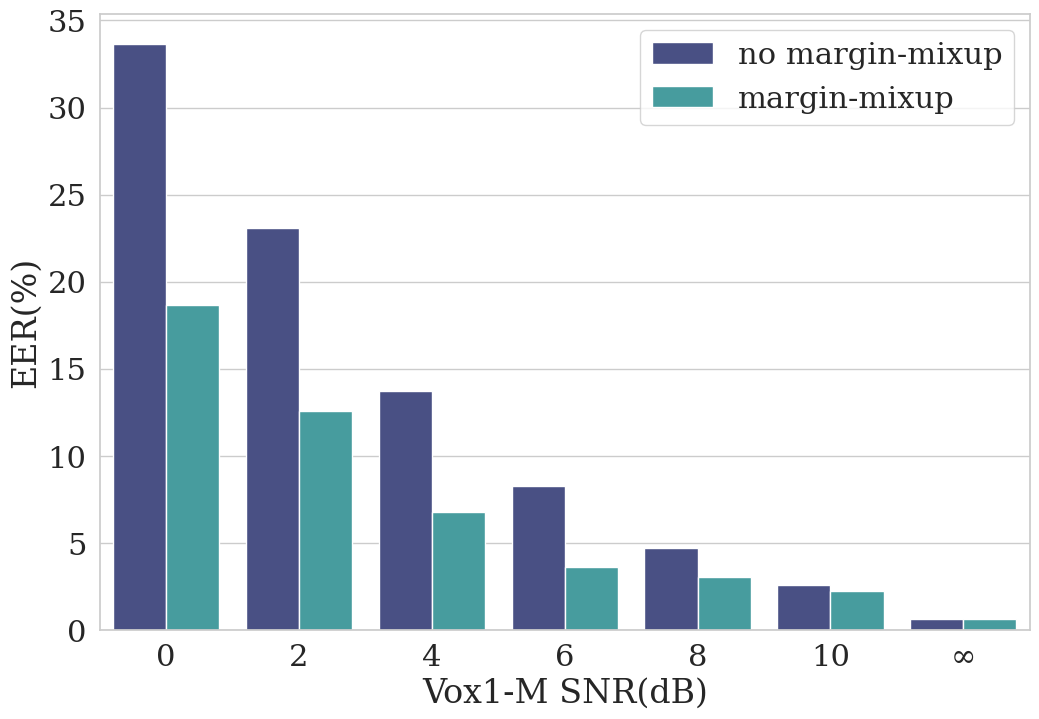}}
\end{minipage}
\caption{Bar chart depicting the impact of the margin-mixup training strategy on the fwSE-ResNet34 model evaluated on the multi-speaker Vox1-M test set with different SNR values.}
\label{fig:snr_vseer}
\end{figure}

Subsequently, we analyze the impact of the $\alpha$ and $\beta$ parameters of the beta distribution described in Section \ref{ssec:margin_mixup} used to sample the interpolation strength $\lambda$ during margin-mixup training in Table \ref{tab:beta_distribution}. We observe a trade-off between the performance on the standard single-speaker test set Vox1-O and the multi-speaker test set Vox1-M at various SNR levels. In the case of a uniform sampling probability of $\lambda$ when $\alpha, \beta = 1$, a significant performance increase is noted on the \textit{0dB} case with a corresponding performance drop on the regular Vox1-O test set. This indicates that more aggressive mixup conditions during training results in more robust embeddings in severe SNR test scenarios, at the cost of regular speaker verification performance. As seen in the case of $\alpha, \beta = 0.2$, no significant degradation is observed on Vox1-O while still attaining more robust embeddings in the multi-speaker scenario.

Finally, to analyze the impact of the SNR values in overlapping speaker verification, we evaluated multiple versions of the Vox1-M test set with a fixed SNR value using the fwSE-ResNet34 baseline model in Figure \ref{fig:snr_vseer}. As expected, the SNR value has a drastic impact on the speaker verification performance with an inverse correlation between SNR and EER. We note that margin-mixup has the most impact in multi-speaker setups with low SNR values. We suspect this is mainly due to the increasing tendency of the baseline models to see the mixed speaker as background noise in higher SNR scenarios.

\section{Conclusion}
\label{sec:conclusion}
In this paper we presented margin-mixup, a training strategy to make speaker embeddings more robust in a multi-speaker audio setup. In contrast to other approaches, margin-mixup requires no architectural changes to speaker verification pipelines to adapt to the multi-speaker scenario, while attaining significant performance improvements in this challenging condition. Training our baseline models with the proposed margin-mixup strategy improves the EER on average with 44.4\% relative over the baseline performance on a multi-speaker version of the original VoxCeleb1 test set. In future work, we aim to extend margin-mixup with a more suitable similarity metric then the cosine distance to increase the multi-speaker modelling capabilities.

%Speaker utterances are mixed at batch-level with a randomly chosen interpolation weight. This interpolation strength is also enforced on the margin penalties applied in our adapted version of the popular AAM-softmax loss function.

\vfill\pagebreak

\bibliographystyle{IEEEbib}
\bibliography{strings,refs}

\end{document}